\pdfoutput=1
\documentclass[pre,aps,twocolumn,showpacs,superscriptaddress]{revtex4-1}
\usepackage{bm}
\usepackage{mathptmx}
\usepackage{enumerate}
\usepackage{epsfig,amsopn}
\usepackage{graphicx}
\usepackage{amsmath,amssymb}
\usepackage{natbib}
\usepackage{braket}
\usepackage{comment}
\usepackage{color}
\def\bea{\begin{eqnarray}}
\def\eea{\end{eqnarray}}

\def\nn{\nonumber}

\newcommand{\eref}[1]{Eq.~(\ref{#1})}%
\newcommand{\Eref}[1]{Equation~(\ref{#1})}%
\newcommand{\fref}[1]{Fig.~\ref{#1}} %
\newcommand{\Fref}[1]{Figure~\ref{#1}}%
\newcommand{\sref}[1]{Sec.~\ref{#1}}%
\newcommand{\aref}[1]{Appendix~\ref{#1}}%

\begin{document}

\title{Driven inelastic Maxwell gases}
\author{V. V. Prasad} 
\affiliation{Raman Research Institute, Bangalore - 560080, India} 
\author{Sanjib Sabhapandit} 
\affiliation{Raman Research Institute, Bangalore - 560080, India} 
\author{Abhishek Dhar} 
\affiliation{International centre for theoretical sciences, TIFR,
Bangalore - 560012, India}

\date{\today}

\begin{abstract}

We consider the inelastic Maxwell model, which consists of a
collection of particles that are characterized by only their
velocities and evolving through binary collisions and external
driving. At any instant, a particle is equally likely to collide with
any of the remaining particles.  The system evolves in continuous time
with mutual collisions and driving taken to be point processes with
rates $\tau_c^{-1}$ and $\tau_w^{-1}$, respectively. The mutual
collisions conserve momentum and are inelastic, with a coefficient of
restitution $r$. The velocity change of a particle with velocity $v$,
due to driving, is taken to be $\Delta v=-(1+r_w) v+\eta$, where
$r_w\in [-1,1]$ and $\eta$ is Gaussian white noise.  For $r_w\in(0,1]$,
this driving mechanism mimics the collision with a randomly moving
wall, where $r_w$ is the coefficient of restitution.  Another special
limit of this driving is the so-called Ornstein-Uhlenbeck process
given by $\frac{dv}{dt}=-\Gamma v+\eta$. We show that while the
equations for the $n$-particle velocity distribution functions
($n=1,2,\dotsc$) do not close, the joint evolution equations of the
variance and the two-particle velocity correlation functions
close. With the exact formula for the variance we find that, for
$r_w\ne-1$, the system goes to a steady state.  Also we obtain the exact tail of the velocity 
distribution in the steady state. On the other hand, for
$r_w=-1$, the system does not have a steady state. Similarly, the
system goes to a steady state for the Ornstein-Uhlenbeck driving with
$\Gamma\not=0$, whereas for the purely diffusive driving ($\Gamma=0$),
the system does not have a steady state.

\end{abstract}


\maketitle

\section{Introduction}
\label{introduction}

A gas of particles undergoing elastic collisions evolves to an
equilibrium state where the single-particle velocity distribution is
Gaussian (Maxwell distribution). For such an isolated system, the
collisions merely distribute energy among the particles while
keeping the total energy constant. In contrast, if the collisions
between particles are inelastic, the system dissipates energy upon
collisions; the change of energy in each binary collision is given by
$\Delta E = - \frac{1}{2}(1-r^2)\bigl[\frac{1}{2} m (\Delta
v)^2\bigr]$, where $r$ is the coefficient of restitution, $m$ is the
mass of the particles, and $\Delta v$ is the relative velocity along
the direction of the collision. It is indeed possible to go from an
inelastic to a quasielastic system of particles within the same
experimental setup using tunable repulsive
interactions~\cite{Merminod:14}.  In a system of inelastic gas
starting from a spatially homogeneous state, the total energy
initially decreases according to the famous Haff's law~\cite{Haff:83}
with $E(t)=E_0 (1+t/t_*)^{-2}$, where $t_*\propto (1-r^2)^{-1}$.  At
long times the particles form high-density
clusters~\cite{goldhirsch:93} with typical mass growing with time as
$M \sim t^\delta$. In this late time regime, the conservation of
momentum dictates that the energy of the system decreases with time as
$E(t)\sim t^{-\delta}$. In one dimension, in this late time regime,
the inelastic gas behaves like a perfectly inelastic sticky gas
(ballistic aggregation model), which can be described by the inviscid
Burgers equation~\cite{Ben-naim:99}. For the sticky gas, scaling
arguments~\cite{Carnevale:90} as well as exact
calculation~\cite{Fracebourg:99} gives $\delta=2/3$. There is no
exact calculation for higher dimensions, and the validity of scaling
arguments as well as the Burgers-like equation is not
clear~\cite{Nie:02, Rajesh:14}.

 In order to keep an inelastic gas in a steady state, it is clearly
necessary to inject energy into the system.  It has been found that,
if energy injection (heating) into the system takes place only at the
boundaries, then clustering of the particles still persist in the bulk
of the system~\cite{kadanoff:95, episov:97, Opsomer:12}, although
there is some evidence of nonclustering for rod-shaped
objects~\cite{Harth:13}.  These studies indicate the need of uniform
heating in order to obtain a spatially homogeneous steady state for
regular granular matter. For such a uniformly driven system of
inelastic particles, one of the most interesting questions is the
velocity distribution in the steady state.  Experiments on driven
granular systems have found non-Gaussian velocity
distributions~\cite{Olafsen:99, Kudrolli:00}. The velocity
distribution found in some of the
experiments \cite{Losert:99,Rouyer:00,Aranson:02} follow the form
$P(v)\sim \exp(-A |v|^\alpha)$ with $\alpha\approx 1.5$. Our interest
in this paper is in the uniformly driven inelastic granular gas.

In analytical studies based on kinetic theory methods, one constructs
the evolution equation for the single-particle velocity distribution
function (ignoring spatial correlations for the 
homogeneous gas). Due to the binary collisions, the equation for the
single-particle distribution depends on the two-particle distribution,
the two-particle distribution depends on the three-particle
distribution, and so on --- creating a hierarchy of equations for the
probability distributions, similar to the BBGKY hierarchy. Usually one
circumvents this problem by invoking the \emph{molecular chaos
hypothesis}, which assumes that the colliding particles are
uncorrelated before a collision, and hence, factorize the two-particle
distribution into two one-particle distributions, resulting in a
closed (\emph{Boltzmann}) equation for the single-particle velocity
distribution. Modeling the uniform heating by adding a diffusive term
in the Boltzmann equation, van Noije and Ernst \cite{Noije:98} have
calculated the steady state velocity distribution. They found a
stretched exponential tail with $\alpha=1.5$, for inelastic hard
sphere gas (where the collision rate is proportional to the relative
velocity of the colliding particles).  On the other hand, the numerical
studies by van Zon and McKintosh \cite{van zon:04} have found a
continuous spectrum of possible exponents ranging up to $\alpha<2$
rather than a universal exponent $\alpha=1.5$.  Intrigued by the
differences in the two results, in this paper we investigate one of
the simplest, yet nontrivial, models of inelastic gases, namely the
Maxwell model.

In the inelastic Maxwell model, introduced by Ben-Naim and
Krapivsky~\cite{Ben-naim:00}, the Boltzmann equation for the
single-particle velocity distribution (assuming product form of
two-point distribution) is further simplified, by taking the rate of
collision to be independent of the velocities of the colliding
particles.  In this case it was shown~\cite{Antal:02,Santos:03} that,
with the diffusive driving, the steady-state velocity distribution has
a form $P(v)\sim \exp(-A|v|)$, while it becomes Gaussian in the
elastic limit~\cite{Barrat:07}.  Recently~\cite{prasad}, we studied a
discrete time version of the inelastic Maxwell model, and showed that
some exact results could be obtained, without taking recourse to the
molecular chaos hypothesis.  In particular, it was observed that the 
equations for the variance and the two-particle correlations of 
the velocities close within themselves \emph{exactly}, even though 
the equations for the velocity distributions have the usual hierarchy.  
From the exact evolution of these equations, we find that \emph{purely
diffusive driving is not enough} to sustain the steady state, as it
causes the variance and the correlations to increase linearly with
time. This simply follows from the fact that the total momentum of the
system also diffuses.  As a result, the assumption of ``molecular
chaos'' is invalid.  We then showed that this problem has a physically
motivated resolution ---namely by introducing a different scheme of
driving.  Wall collisions of vibrated particles do not conserve total
momentum and, is the typical way of driving in real systems ---we
incorporate this into the driving forces and studied the resulting
steady state.  Importantly, we were also able to obtain the exact
tails for the velocity distributions in the steady state.

In this paper, we extend the results of discrete time dynamics to a
system evolving in continuous time.  For the case of continuous time
dynamics, we again illustrate that the evolution equations for the
variance and the two-point correlations form a closed set, even though
the equations for the distribution functions themselves do not close.
An exact mapping to the discrete model enables us to obtain the
high-energy tail of the velocity distribution for the continuous time
model.  We also find that the Ornstein-Uhlenbeck driving is a special
case of our model. This makes it possible to obtain the exact tail
behavior of the velocity distribution in a steady state when driven by
an Ornstein-Uhlenbeck process. Thus our work compliments the previous
studies~\cite{Marconi:02, puglisi:98, Costantini:07} where the
probability density function (PDF) of the velocity is calculated as a
series expansion around the Maxwellian.  The exact coupled equations, for
the variance and correlation, permits one to predict the existence of
steady states in different parameter regimes of the system.  In
particular, we show the absence of a steady state for a continuous
time system driven by purely diffusive driving, which is a special
case of the Ornstein-Uhlenbeck process.

The continuous time model, introduced here has another significance in
connection with real systems. In experimental studies of driven
granular systems, the driving is caused by the collisions of the
particles with the vibrating walls of the container. Like
interparticle collisions, the wall-collisions also occur as a point
process in time, with finite change in particle velocities. In
contrast, the typical analytical models employ continuous driving
schemes like diffusive or Ornstein-Uhlenbeck processes. We propose that
the model that is introduced here is a better scheme of driving in the
sense that the driving is taken to be a point process in time with a
rate associated with it.

The rest of the paper is organized as follows. We first define the
rules for the inelastic collision between a pair of particles as well
as the driving mechanism in \sref{rules}. Next, in 
\sref{continuous point process}, we discuss
the Maxwell gas with continuous time dynamics with both collision and
driving occurring as Poisson processes. We find an exact formula for
the coupled evolution for the variance and the two-particle
correlation function. We also obtain the exact tail of the steady-state velocity distribution in the thermodynamic limit of large number
of particles. We point out the correspondence between this continuous
time model and the discrete model discussed in~\aref{discrete}.
In~\sref{OU asymptotic}, we take a particular limit of the driving
parameters to obtain the Ornstein-Uhlenbeck process.  The absence of
steady state for a system with diffusive driving is easily obtained
from the evolution equations. Finally, we conclude
in \sref{conclusion}.  The Maxwell model evolving with discrete
dynamics and some of the details are given in the appendix.

\section{Collision rule and driving mechanism}
\label{rules}

For simplicity, we assume the velocities of the particles to be
single component (one dimensional). In the inelastic collisions
between two particles (say, $i$ and $j$), their velocities are 
modified from $(v_i^*,v_j^*)$ to $(v_i, v_j)$ according to
$(v_i-v_j)=-r(v^*_i-v^*_j)$ while keeping the total momentum unchanged
$v_i+v_j= v_i^* + v_j^*$, where $r$ is the coefficient of restitution
and we have set the masses of the particles to unity. Combining the
above two rules, one gets the postcollision velocities in terms of
the precollision velocities as
\begin{subequations}
\label{interparticle collision}
\begin{align}
v_{i} &= \frac{(1-r)}{2} v_i^* + \frac{(1+r)}{2} v_{j}^*,\\ v_{j}
&= \frac{(1+r)}{2} v_i^* + \frac{(1-r)}{2} v_{j}^*.
\end{align}
\end{subequations}

Our model of driving is inspired by the collision of particles with a
vibrating wall, where the post-collision velocity $v_i$ of a particle
is related to its precollision velocity $v_i^*$ by
$(v_i-V_w)=-r_w(v^*_i-V^*_w)$, with $r_w$ being the coefficient of
restitution between the wall and particle collision. However, the
velocity of a massive wall remains unchanged during a collision,
$V^*_w=V_w$. Therefore, one has $v_i=-r_w v^*_i + (1+r_w) V_w$. One
can further assume that, the velocity of the wall in each collision is an  uncorrelated random variable.  Therefore, in our model of driving,
the velocity of a particle is modified according to
\begin{equation}
v_i=-r_w v_i^* + \eta,
\label{wall collision}
\end{equation}
where $\eta$ is a Gaussian random variable with zero mean and variance
$\sigma^2$, drawn independently at each time.

For physical collisions, the coefficients of restitution $\{r,
r_w\}\in [0,1]$, where $\{r, r_w\}=1$ corresponds to the elastic
collision, whereas $\{r, r_w\}=0$ corresponds to the sticky collision.
However, it is important to note that, as a mathematical model of a
driven dissipative system, Eqs.~\eqref{interparticle collision}
and \eqref{wall collision} are well defined over the entire range
$\{r,r_w\} \in [-1,1]$. Therefore, we investigate this model over this
entire range and treat $r_w$ and $\sigma$ as independent parameters.

\section{The Maxwell model}
\label{continuous point process}

The model consists of a set of $N$ identical particles characterized
by only their one-component velocities $v_i$, with $i=1,2,...,N$. The
initial velocities are taken independently from a Gaussian
distribution. There is no spatial structure in the model.  The system
evolves in continuous time and we consider both the interparticle
collisions and the driving to be uncorrelated random processes in time
(see \aref{discrete} for the model with discrete time dynamics).  We
let the particles of each pair collide at a rate $g\tau_c^{-1}$,
according the the collision rule given by \eref{interparticle
collision}. On the other hand, each particle is driven at a rate
$g\tau_w^{-1}$, according to the driving mechanism given by \eref{wall
collision}.

Let us define a set of distribution functions for the system,
\begin{subequations}
\begin{align}
F_1(u_1,t)&\equiv\sum\limits_{i=1}^{N}
\langle\delta(u_1-v_i(t))\rangle,  \\
F_2(u_1,u_2,t)&\equiv\sum\limits_{i=1}^{N}\displaystyle\sum\limits_{j\neq
i}^{N}\langle \delta(u_1-v_i(t))\delta(u_2-v_j(t))\rangle, \\
F_3(u_1,u_2,u_3,t)&\equiv\sum\limits_{i=1}^{N}\sum\limits_{j\neq
i}^{N}\sum\limits_{k\neq
i,j}^{N}\langle \delta(u_1-v_i(t))\delta(u_2-v_j(t))\nn\\
&\hspace{3cm}\times\delta(u_3-v_k(t))\rangle, 
\end{align}
\end{subequations}
and so on.  The evolution equations of the above distributions form a
hierarchy, and the first two such equations are given by
\begin{widetext}
\begin{subequations}
\label{F-eqn}
\begin{align}
\label{engy_wall_collision_point_process}
&\frac{\partial}{\partial
t}F_1(v_1,t)=g\tau_c^{-1}\left[\int
dv_2\overline{T}(v_1,v_2)F_2(v_1,v_2,t)\right]+g\tau_w^{-1}\left[\int
dv_1^*F_1(v_1^*,t)\langle\delta\left(v_1-[-r_wv_1^*+\eta_1]\right)\rangle_{\eta_1}-F_1(v_1,t)\right], \\
\label{corrlsn_constantini collision_point_process}
&\frac{\partial}{\partial t}F_2(v_1,v_2,t)=g\tau_c^{-1}\left[\overline{T}(v_1,v_2)F_2(v_1,v_2,t)+\int dv_3\left[\overline{T}(v_1,v_3)+\overline{T}(v_2,v_3)\right]F_3(v_1,v_2,v_3,t)\right]\nn\\
&\quad+g\tau_w^{-1}\left[\int dv_1^*F_2(v_1^*,v_2,t)\langle\delta\left(v_1-[-r_wv_1^*+\eta_1]\right)\rangle_{\eta_1}+
\int
dv_2^*F_2(v_1,v_2^*,t)\langle\delta\left(v_2-[-r_wv_2^*+\eta_2]\right)\rangle_{\eta_2}-2F_2(v_1,v_2,t)\right].
\end{align}
\end{subequations}
\end{widetext}
The  first square bracket in the right-hand side of
Eqs.~\eqref{engy_wall_collision_point_process}
and \eqref{corrlsn_constantini collision_point_process} gives the
contribution from the interparticle collisions, with
$\overline{T}(v_i,v_j)$ defined as, $\overline{T}(v_i,v_j)S(v_i,v_j)=
r^{-1}S(v_i^*,v_j^*)-S(v_i,v_j)$.  The operator $\overline{T}$ acts
only on the two variables designated by the arguments of the
operator. The second set of square brackets in
Eqs.~\eqref{engy_wall_collision_point_process}
and \eqref{corrlsn_constantini collision_point_process} are the
contribution from the driving, where the angular brackets refer to the
averaging over the noise distribution. Various approximation schemes
have been used in the past to break the hierarchy of similar
equations~\cite{Brey:04, Costantini:07}. In the following, we show
that exact closed set of coupled equations can be obtained for the
variance and the two-particle correlation function, whose solution, in
turn, can be used to close the hierarchy for the single-particle
distribution function in the $N\to\infty$ limit.

The variance and the two-particle correlation function  
can be obtained using the above-defined distributions
as
\begin{subequations}
\begin{align}
\Sigma_1(t)&=\frac{1}{N}\int dv_1 v_1^2 F_1(v_1,t), \\
\Sigma_2(t)&=\frac{1}{N(N-1)}\int dv_1 dv_2 v_1
  v_2 F_2(v_1,v_2,t). 
\end{align}
\end{subequations}
Now, multiplying \eref{engy_wall_collision_point_process} by $v_1^2$
and then integrating over $v_1$, and multiplying
\eref{corrlsn_constantini collision_point_process} by $v_1 v_2$ and integrating
over both $v_1$ and $v_2$, yield a closed set of equations for
$X(t)=[\Sigma_1(t), \Sigma_2(t)]^T$, given by
\begin{equation}
\frac{d X(t)}{d t} = g\bigl[{\bf R} X(t) + C\bigr],
\label{continuum matrix abstract finite gamma}
\end{equation}
where 
\begin{equation} 
{\bf R}=\left[\begin{array}{cc} -\left(\frac{(1-r^2)(N-1)}{2\tau_c}
+\frac{1-r_w^2}{\tau_w}\right) & \frac{(1-r^2)(N-1)}{2\tau_c}~\\ \\
\frac{(1-r^2)}{2\tau_c} & -\left(\frac{(1-r^2)}{2\tau_c} 
+\frac{2(1+r_w)}{\tau_w}\right) \end{array}\right], 
\label{collision wall dissipation point process}
\end{equation}
and $C=[\tau_w^{-1}\sigma^2, 0]^T$. Note that \eref{continuum matrix
abstract finite gamma} is exact and no approximation is made in
arriving at it from \eref{F-eqn}.

\begin{figure}
\includegraphics[width=.9\hsize]{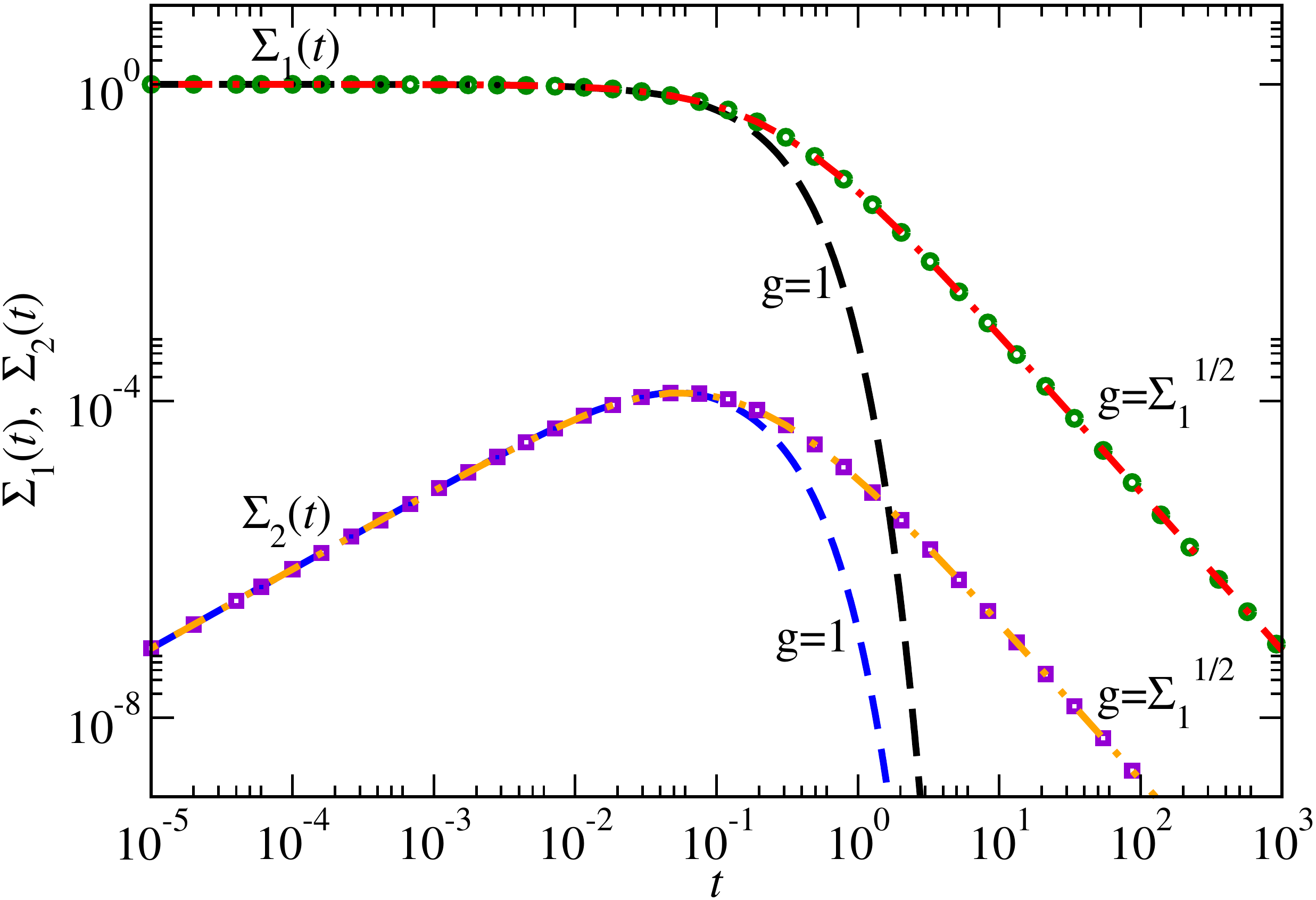}
\caption{\label{Figure2} The variance $\Sigma_1(t)$ and the
two-particle correlation $\Sigma_2(t)$ of the velocities for a
cooling inelastic gas with $1000$ particles with  $r=1/2$, 
in the absence of the driving (static walls) 
for the two cases: (a) The rate of collision is independent of the
variance $g=1$, and (b) the rate of collision is proportional to the
variance $g=\Sigma_1^{1/2}(t)$. For $g=1$, the lines plot the exact
analytical expressions given by \eref{eq-1}. For $g=\Sigma_1^{1/2}(t)$, the
lines plot the approximate expressions given by Eqs.~\eqref{eq3}
and \eqref{eq4}, while the points are obtained by exact numerical
evaluation of the equation \eref{eq0}.}
\end{figure}

Now, in the Maxwell model with the collision rates proportional to the
typical velocity~\cite{Ben-naim:02}, one uses
$g=\Sigma_1^{1/2}$. However, it makes \eref{continuum matrix abstract
finite gamma} nonlinear, and hence, the analysis becomes difficult.
On the other hand, it is clear from both \eref{F-eqn}
and \eref{continuum matrix abstract finite gamma} that the steady-state  properties are independent of the choice of $g$. Therefore, we
set $g=1$ as in Ref.~\cite{Ben-naim:00}, which makes \eref{continuum matrix
abstract finite gamma} linear. This would, of course, change the time-dependent  properties.  For example, in the absence of the driving
($\sigma=0$), the two cases, $g=\Sigma_1^{1/2}$ and $g=1$, yield
different cooling laws, as discussed in \aref{HCS} and shown
in \fref{Figure2}.

In the presence of the driving ($\sigma\not = 0$), one
again expects the approach to the steady state to differ for the
two choices of $g$. We analyze \eref{continuum matrix abstract finite
gamma} for the particular choice of $g=1$.  In this case, the linear
equation can be exactly solved. 
The variance and the two-particle correlation are
given by
\begin{subequations}
\label{Sigma-general}
\begin{align}
&\Sigma_1(t)=\frac{\Sigma_1(0)}{\lambda_+-\lambda_-}
\Bigl[ (R_{22} -\lambda_-)
e^{-\lambda_- t} + (\lambda_+-R_{22}) \, e^{-\lambda_+ t}  \Bigr]\notag \\
&+\frac{\tau_w^{-1}\sigma^2}{\lambda_+-\lambda_-}
\biggl[
\frac{R_{22}-\lambda_-}{\lambda_-}\bigl(1-e^{-\lambda_- t}\bigr) 
+\frac{\lambda_+ -R_{22}}{\lambda_+}\bigl(1-e^{-\lambda_+
t}\bigr)
 \biggr], 
\intertext{and}
&\Sigma_2(t)=\frac{\Sigma_1(0)R_{21}}{\lambda_+-\lambda_-}
\Bigl[e^{-\lambda_- t} -e^{-\lambda_+ t}\Bigr]\notag\\
&\qquad+\frac{\tau_w^{-1} \sigma^2 R_{21}}{\lambda_+-\lambda_-}
\biggl[\frac{1}{\lambda_-} (1-e^{-\lambda_- t}) 
-\frac{1}{\lambda_+}(1-e^{-\lambda_+t})\biggr] , 
\end{align}
\end{subequations} 
respectively, where $-\lambda_\pm$ are the eigenvalues of
$\mathbf{R}$, given by \eref{lambda-pm},  and $R_{ij}=|\mathbf{R}_{ij}|$.

\begin{figure}
\includegraphics[width=.9\hsize]{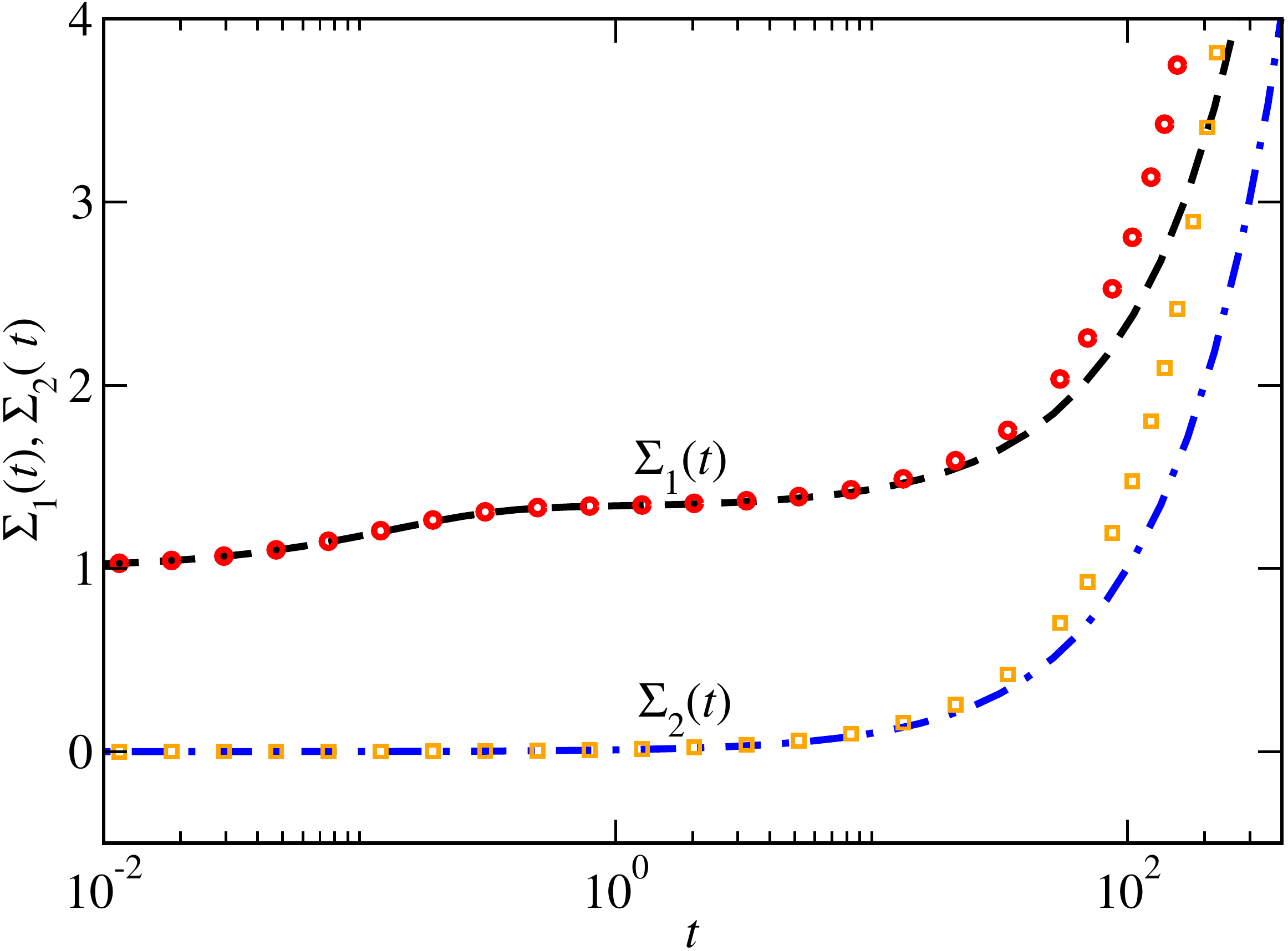}
\caption{\label{Figure3} The variance $\Sigma_1(t)$ and the
two-particle correlation $\Sigma_2(t)$ of the velocities, for an
inelastic gas with $1000$ particles with $r=1/2$ driven by wall
collisions with $\sigma=1$ and $r_w=-1$ for the two cases: (a) The
rate of collision is independent of the variance $g=1$ and (b) the
rate of collision is proportional to the variance
$g=\Sigma_1^{1/2}(t)$. For $g=1$, the lines plot the exact analytical
expressions given by \eref{sigma for rw=-1}.  For
$g=\Sigma_1^{1/2}(t)$, the points are obtained by exact numerical
evaluation of the equation \eref{continuum matrix abstract finite
gamma}. }
\end{figure}

Now, for the case $r_w=-1$, one of the eigenvalues of ${\bf R}$
becomes zero ($\lambda_-=0$), while the other is negative
($\lambda_+=R_{11} + R_{22} >0$). For these particular values of
$\lambda_\pm$, the above expressions become
\begin{subequations}
\label{sigma for rw=-1}
\begin{align}
\Sigma_1(t)=&\frac{\Sigma_1(0)}{\lambda_+}
\Bigl[R_{22} + R_{11}\, e^{-\lambda_+ t} \Bigr]
 + \frac{\sigma^2}{\tau_w} \frac{R_{11}}{\lambda_+^2}
\Bigl[1-e^{-\lambda_+ t} \Bigr] \notag\\
&+ \frac{\sigma^2}{\tau_w} \frac{R_{22}}{\lambda_+}\, t, \\
\Sigma_2(t)=&
\frac{\Sigma_1(0)R_{21}}{\lambda_+}
\Bigl[1 -e^{-\lambda_+ t}\Bigr]
 - \frac{\sigma^2}{\tau_w} \frac{R_{21}}{\lambda_+^2}
\Bigl[1-e^{-\lambda_+ t} \Bigr] 
\notag\\
&+ \frac{\sigma^2}{\tau_w} \frac{R_{21}}{\lambda_+}\, t.
\end{align}
\end{subequations}
Thus, both $\Sigma_1(t)$, and $\Sigma_2(t)$ eventually increase
linearly with time and the system does not have a steady state for
$r_w=-1$ when the driving is present ($\sigma\not= 0$), which is shown in \fref{Figure3}.

\begin{figure}
\includegraphics[width=.9\hsize]{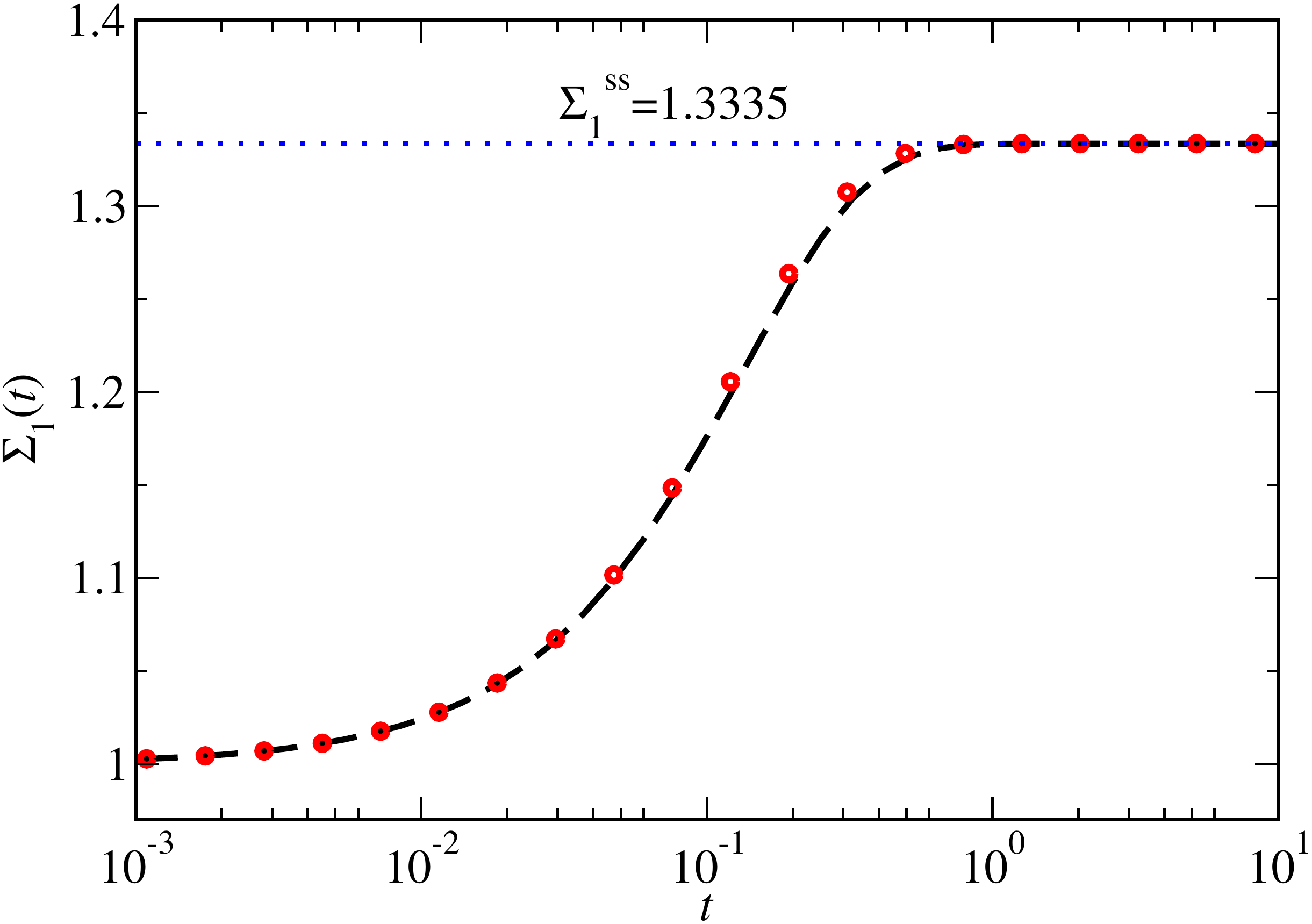}
\caption{\label{Figure4}The variance $\Sigma_1(t)$ of the velocities, for an
 inelastic gas with $1000$ particles with $r=1/2$ driven by wall
collisions with $\sigma=1$ and $r_w=1$ for the two cases: (a) The rate
of collision is independent of the variance $g=1$ and (b) the rate of
collision is proportional to the variance $g=\Sigma_1^{1/2}(t)$. For
$g=1$, the line plots the exact analytical expressions given
by \eref{Sigma-general}.  For $g=\Sigma_1^{1/2}(t)$, the points are
obtained by exact numerical evaluation of the equation \eref{continuum
matrix abstract finite gamma}. The dotted line highlights the steady-state  
value calculated from Eq. (\ref{steady state sigma1 sigma2}a). }
\end{figure}

On the other hand, for $-1 < r_w \le 1$, since both the eigenvalues of
${\bf R}$ are negative ($\lambda_\pm >0$) (see \aref{HCS}), the system
reaches a steady state as shown in \fref{Figure4}. The steady state values of $\Sigma_1$ and
$\Sigma_2$ can be obtained by either taking the limit of $t\to \infty$
in \eref{Sigma-general} or by setting the left-hand side
of \eref{continuum matrix abstract finite gamma} to zero. From the
latter, it is clear that the steady-state values are independent of
the choice of $g$.  We denote the steady-state values of the variance
and the two-particle correlation by $\Sigma_1^\mathrm{ss}$ and
$\Sigma_2^\mathrm{ss}$, respectively, and they are given by
\begin{widetext}
\begin{subequations}
\label{steady state sigma1 sigma2}
\begin{align}
\Sigma_1^\mathrm{ss}=
\frac{\sigma^2\bigl[(1-r^2)+4(1+r_w)(\tau_c/\tau_w)\bigr]}
{(1-r_w^2)(1-r^2) + 2 (1+r_w) \bigl[(1-r^2) (N-1)+ 2(1-r^2_w)
(\tau_c/\tau_w)\bigr]}~, \\
\Sigma_2^\mathrm{ss}=
\frac{\sigma^2(1-r^2)}
{(1-r_w^2)(1-r^2) + 2 (1+r_w) \bigl[(1-r^2) (N-1)+ 2(1-r^2_w)
(\tau_c/\tau_w)\bigr]}~.
\end{align}
\end{subequations}
\end{widetext}

We now analyze the above expressions for various relative rates
$\tau_c/\tau_w$. Let us take $\tau_c/\tau_w\sim N^\xi$ for large $N$,
where $\xi$ is a real number.  From \eref{steady state sigma1 sigma2},
we observe that for $\xi<0$, both $\Sigma_1^\mathrm{ss} $ and
$\Sigma_2^\mathrm{ss} $ vanish as $O(1/N)$ for large $N$.  Similarly,
for $0<\xi<1$, they again vanish as $\Sigma_1^\mathrm{ss} \sim
O(1/N^{1-\xi})$ and $\Sigma_2^\mathrm{ss} \sim O(1/N)$ for large $N$.
Only for $\xi\ge1$ do we get a nonzero steady-state variance for large
$N$, given by
\begin{equation}
\Sigma_1^\mathrm{ss}=
\frac{2\sigma^2 (\tau_c/\tau_w)}{N(1-r^2)+2(1-r_w^2)(\tau_c/\tau_w)},
\label{steady state sigma1 large N}
\end{equation}
whereas the two-particle correlation function vanishes as
$\Sigma^\mathrm{ss}_2 \sim O(1/N^{\xi})$.

Due to the mean-field nature of the problem, it is reasonable to
assume that the  rate, $\tau_c^{-1}$, of interparticle collisions 
is inversely proportional to the total number of pairs [$\tau_c \propto N (N-1)/2$] whereas  
the rate, $\tau_w^{-1}$, of 
 driving is inversely proportional to the total number of particles
 ($\tau_w \propto N$). This is analogous to taking the coupling constant
proportional to $1/N$ in infinite-ranged spin models.  Indeed, if we set
$\tau_c/\tau_w = \gamma (N-1)$, then \eref{steady state sigma1 sigma2}
becomes identical to those obtained for the discrete time dynamics
[see \eref{steady state discrete}]. In particular, in the limit
$N\to \infty$, the steady-state variance becomes independent of $N$, 
\begin{equation}
\Sigma_1^\mathrm{ss}=\frac{2\gamma\sigma^2}{(1-r^2)+2\gamma(1-r_w^2)}.
\end{equation}
 Moreover, since the two-particle correlation function vanishes in the
limit of large $N$, we can factorize the multiparticle distribution
functions in terms of the single-particle distribution function
in \eref{F-eqn}, e.g., $F_2(v_1,v_2)= F_1(v_1) F_1(v_2)$. Therefore,
in the steady state [the time derivatives in \eref{F-eqn} are zero],
multiplying
\eref{engy_wall_collision_point_process} by  $\exp(-\lambda v_1)$  and
then integrating over $v_1$ we obtain the equation satisfied by the
generating function $Z(\lambda)$ as
\begin{equation}
Z(\lambda)=q Z(\epsilon\lambda)Z([1-\epsilon]\lambda) + (1-q)
Z(r_w \lambda)f(\lambda),
\label{Z-continuous}
\end{equation}
 where $q=1/(1+\gamma)$ and
$f(\lambda)=\exp(\lambda^2\sigma^2/2)$. Since the velocity
distribution is even, we have $Z(-\lambda)=Z(\lambda)$.  The above
equation is identical to \eref{recursion_mgf} obtained for the
discrete time dynamics. Therefore, as expected, both the continuous
time and the discrete time dynamics yield the same steady state.


For the particular case $r_w=1$, we can obtain $Z(\lambda)$ as an
infinite product involving simple poles by iteratively solving
\begin{math}
Z(\lambda)=[1-(1-q) f(\lambda)]^{-1}\, q
Z(\epsilon\lambda)Z([1-\epsilon]\lambda) . 
\end{math}
Therefore, the tail of the velocity distribution is exponential
$P(v)\sim \exp(-|v|/v^*)$, where $v^*$ is determined by the pole
closest to the origin, coming from the prefactor $[1-(1-q)
f(\lambda)]^{-1}$.

\begin{figure}
\includegraphics[width=0.95\hsize]{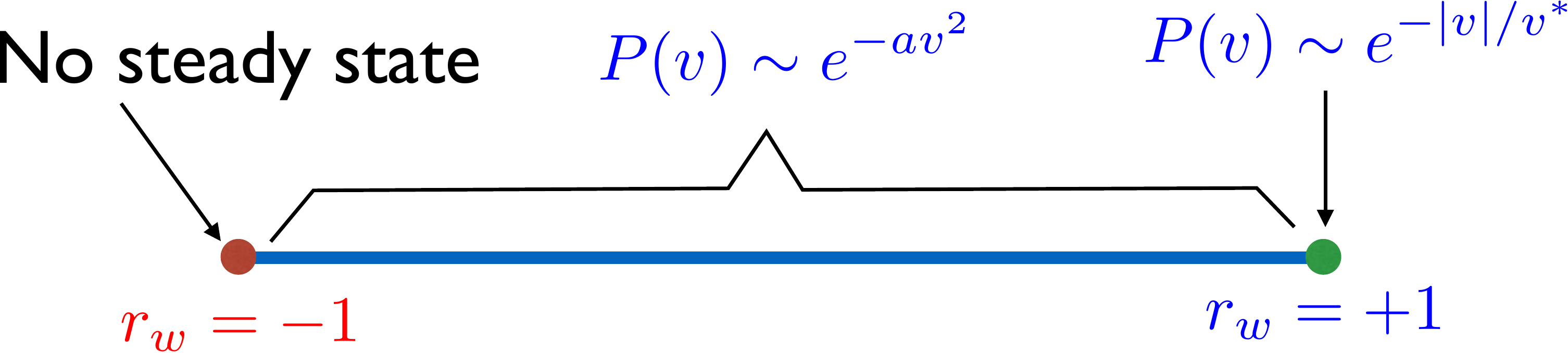}
\caption{\label{Figure1} The summary of  the results for the
PDF of the velocity distribution for different cases of $r_w \in
[-1,1]$. For $r_w=-1$, the system does not reach a steady state, and
the average energy and the two-particle correlation eventually
increases linearly with the time.  For $r_w=1$, the steady-state PDF
has an exponential tail, whereas for $-1 < r_w < 1$, the tail of the
PDF for very large velocities is Gaussian.}
\end{figure}

On the other hand, for $|r_w| < 1$, if we assume the form
$P(v)\sim \exp(-A \mid v\mid^{\alpha})$ for the PDF, then for $\alpha
>1$, the function $Z(\lambda)$ is analytic.  If $Z(\lambda)$ is known,
then the large deviation tail of the velocity distribution can be
obtained by the saddle-point approximation,
\begin{equation}
P(v)\approx\frac{\exp\bigl[\mu(\lambda^*) +\lambda^*
v\bigr]}{\sqrt{2\pi|\mu''(\lambda^*)|}}, 
\end{equation}
where $\mu(\lambda)=\ln
Z(\lambda)$ and the saddle point $\lambda^*(v)$ is implicitly given by
the equation $\mu'(\lambda^*)=-v$.  Now, if near the saddle point
$\mu(\lambda) \sim b|\lambda|^\beta$, one finds
\begin{equation}
\lambda^*=-\mathrm{sign}(v)
\left[\frac{|v|}{(b\beta)}\right]^{1/(\beta-1)}.
\end{equation}   
As a result, $P(v)\sim 
\exp(-A |v|^{\alpha})$,
where $\alpha=\beta/(\beta-1)$ and $A=b(\beta-1)(b\beta)^{-\alpha}$.
Therefore, we substitute the ansatz
$Z(\lambda) \sim \exp(b|\lambda|^\beta)$ with $\beta > 1$
in \eref{Z-continuous}. Since $\epsilon^\beta +(1-\epsilon)^\beta <
1$ for $\epsilon \in (0,1)$ and $\beta >1$, the first term on the
right-hand side of \eref{Z-continuous} becomes negligible compared to
the left-hand side for large $|\lambda| \sim
|v|^{1/(\beta-1)}$. Thus, comparing the exponent of the left-hand side
to that of the second term on the right-hand side, we get $\beta=2$
and $b=(\sigma^2/2) (1-r_w^2)^{-1}$. This implies the Gaussian tail
\begin{equation}
P(v)\approx \sqrt{\frac{1-r_w^2}{2\pi \sigma^2}}
\,\exp\left[-\frac{v^2}{2\sigma^2}
(1-r_w^2)\right].
\label{LD tail}
\end{equation} 
We have verified this result in~ Ref.~\cite{prasad} for the case of discrete
time dynamics. \Fref{Figure1} summarizes the results for different
case of $r_w$.

\section{Ornstein-Uhlenbeck driving}
\label{OU asymptotic}

We now show that the driving mechanism introduced above becomes an
Ornstein-Uhlenbeck process in a special limit. Let us, for the time
being, ignore the interparticle collisions and also set $g=1$. Then
\eref{engy_wall_collision_point_process} becomes
\begin{align}
\label{engy_wall_collision_point_process tw-1 only}
\frac{\partial F_1(v_1,t)}{\partial t} =
\tau_w^{-1}\biggl[\int dv_1^*F_1(v_1^*,t) 
\langle\delta\bigl(v_1 &-[-r_wv_1^*+\eta_1]\bigr)\rangle_{\eta_1}\nn\\ 
&-F_1(v_1,t)\biggr].
\end{align}
In terms of the characteristic function,
$\widetilde{F_1}(k_1,t)\equiv\int dv_1 F_1(v_1,t)\,e^{-ik_1v_1}$, 
the above equation can be written as
\begin{equation}
\label{engy_wall_collision_point_process_tc-1 very less to tw-1 charac 2}
\frac{\partial \widetilde{F_1}(k_1,t)}{\partial t}=\tau_w^{-1}\left[\widetilde{F_1}(-k_1r_w,t)\,e^{-k^2\sigma^2/2}-\widetilde{F_1}(k_1,t)\right].
\end{equation}
The term $\exp(-k^2\sigma^2/2)$ is the characteristic function for a
Gaussian noise with variance $\sigma^2$.  We now consider the limiting
case when $\tau_w\to0$, $ \epsilon_{w}=(1+r_w)\to0$ and
$\sigma^2\to0$, while keeping appropriate ratios fixed. Replacing 
$r_w$ with $-(1-\epsilon_w)$
in \eref{engy_wall_collision_point_process_tc-1 very less to tw-1
charac 2}, and expanding and keeping only up to the lowest-order terms in
the small parameters, we obtain
\begin{equation}
\label{engy_wall_collision_point_process_tc-1 very less to tw-1 charac 3}
\frac{\partial \widetilde{F_1}(k_1,t)}{\partial
t}=\tau_w^{-1}\left[-\epsilon_{w}k_1\frac{\partial \widetilde{F_1}(k_1,t)}{\partial k_1}-\frac{\sigma^2k_1^2}{2}\widetilde{F_1}(k_1,t)\right]
\end{equation}
Therefore, in the limit $\tau_w\to0,~ \epsilon_{w}\to0$, and
$\sigma^2\to0$, while keeping
\begin{equation}
\Gamma=\lim_{\substack{\tau_w \to 0\\\epsilon_{w} \to 0}}  
\frac{\epsilon_w}{\tau_w} \quad\text{and}\quad
D=\lim_{\substack{\tau_w \to 0\\ \sigma^2 \to
0}}  \frac{\sigma^2}{2\tau_w}
\label{tw limit gamma and d}
\end{equation}
fixed, \eref{engy_wall_collision_point_process_tc-1 very less to tw-1
charac 3} becomes
\begin{equation}
\label{engy_wall_collision_point_process_tc-1 very less to tw-1 charac 4}
\frac{\partial \widetilde{F_1}(k_1,t)}{\partial t}
=-\Gamma k_1\frac{\partial \widetilde{F_1}(k_1,t)}{\partial
k_1}-Dk_1^2\widetilde{F_1}(k_1,t).
\end{equation}
This is nothing but the Fokker-Planck equation of an
Ornstein-Uhlenbeck process in the Fourier space, which, in the velocity
space, is given by
\begin{equation}
\label{engy_wall_collision_point_process_tc-1 very less to tw-1 charac 4}
\frac{\partial {F_1}(v_1,t)}{\partial t}
=\Gamma \frac{\partial}{\partial v_1}   \bigl[v_1 F_1\bigr]
+D\frac{\partial^2 F_1}{\partial v_1^2}. 
\end{equation}
 
Thus, in the limit given by \eref{tw limit gamma and d}, our model of
wall driving becomes an Ornstein-Uhlenbeck process, with the
parameters defined as in \eref{tw limit gamma and d}. The matrix ${\bf
R}$ in this case becomes
\begin{equation} 
{\bf R}=\left[\begin{array}{cc} 
 -\left(\frac{(1-r^2)(N-1)}{2\tau_c} +2\Gamma\right) & \frac{(1-r^2)(N-1)}{2\tau_c}~\\ \\
\frac{(1-r^2)}{2\tau_c} & -\left(\frac{(1-r^2)}{2\tau_c}
+2\Gamma\right) \end{array}\right], 
\label{continuum nonzero gamma} 
\end{equation}
and $C=[2D,0]^T$.  The eigenvalues of {\bf R} are given by $-2\Gamma$
and $-2\Gamma - N(1-r^2)/(2\tau_c)$, with the corresponding
eigenvectors $[1,1]^T$ and $[1,-1/(N-1)]^T$, respectively.  For $g=1$,
we can solve \eref{continuum matrix abstract finite gamma} easily by
diagonalizing ${\bf R}$.  This results in two decoupled equations for
the elements of $[y_1(t),y_2(t)]^{T}={\bf S^{-1}}X$, where
\begin{equation} 
{\bf S}=\left[\begin{array}{cc} 
 1 & 1~\\ \\
1 & -\frac{1}{(N-1)}\end{array}\right]~
\rm{and ~ } ~
{\bf S^{-1}}=\frac{N-1}{N}\left[\begin{array}{cc} 
 \frac{1}{N-1} & 1~\\ \\
1 & -1\end{array}\right]~,
\end{equation}
and $\bf {S^{-1} R S}$ is a
diagonal matrix with the eigenvalues of ${\bf R}$ being the diagonal
elements. It is straightforward to find the solutions as
\begin{subequations}
\label{y1 y2 gamma soln}
\begin{align}
&y_1(t)=y_1(0)\exp\left(-2\Gamma
t\right)+\frac{D}{N\Gamma}\left[1-\exp(-2\Gamma t)\right],\\
\intertext{and}
&y_2(t)=y_2(0)\exp\left(-\left[\frac{N(1-r^2)}{2\tau_c}+2\Gamma\right]t\right)
\nn\\
&+\frac{(N-1)4D\tau_c}{N[N(1-r^2)+4\Gamma\tau_c]}\left[1-\exp\left(-\left[\frac{N(1-r^2)}{2\tau_c}+2\Gamma\right]t\right)\right].
\end{align}
\end{subequations}
The initial values, $y_1(0)$ and $y_2(0)$, are obtained in terms of
$\Sigma_1(0)$ and $\Sigma_2(0)=0$. Finally, $\Sigma_1(t)$ and
$\Sigma_2(t)$ can be obtained by using $X={\bf S}[y_1(t),y_2(t)]^{T}$.

Thus, for any nonzero values of $\Gamma$, we see from \eref{y1 y2
gamma soln} that as $t\rightarrow\infty$, both $y_1(t)$ and $y_2(t)$,
and hence $\Sigma_1(t)$ and $\Sigma_2(t)$ approach steady-state
values. They are given by
\begin{subequations}
\begin{align}
\label{asymptotic exact gamma non-zero}
\lim_{t\to\infty}~\Sigma_1(t)&=\frac{D}{\Gamma}\left(\frac{1-r^2+4\Gamma\tau_c}{N(1-r^2)+4\Gamma\tau_c}\right), \\ 
\lim_{t\to\infty}~\Sigma_2(t)&=\frac{D}{\Gamma}\left(\frac{1-r^2}{N(1-r^2)+4\Gamma\tau_c}\right).
\end{align}
\end{subequations}
These can be also obtained from \eref{steady state sigma1 sigma2} by
taking the limits given by \eref{tw limit gamma and d}.

Let us consider the special case, where the dissipative term
$\Gamma=0$. Here the driving is modeled by a Weiner process (diffusive
driving), $dv_i/dt=\sqrt{2D}\,\eta_i $. In this case, one of the
eigenvalues of ${\bf R}$ becomes zero, while the other is
$-N(1-r^2)/(2\tau_c)$.  The zero eigenvalue indicates a nonstationary
state. The exact solution in the diagonal basis is given by
\begin{subequations}
\begin{align}
\label{y1_2 gamma0 soln}
y_1(t)&=y_1(0)+\frac{2Dt}{N},\\
y_2(t)&=y_2(0)\exp\left(-\frac{N(1-r^2)t}{2\tau_c}\right)\hspace{2cm}\nn\\
&+\frac{(N-1)4D\tau_c}{N^2(1-r^2)}\left[1-\exp\left(-\frac{N(1-r^2)t}{2\tau_c}\right)\right]. 
\end{align}
\end{subequations}
We can obtain $\Sigma_1(t)$ and $\Sigma_2(t)$ exactly for any time
$t$ by inverting $y_1(t)$ and $y_2(t)$. There asymptotic forms for
large $t$ are given by
\begin{subequations}
\begin{align}
\label{asymptotic var}
\Sigma_1(t)&=\frac{\Sigma_1(0)}{N}+\frac{(N-1)4D\tau_c}{N^2(1-r^2)}
+\frac{2D}{N}t~, 
\\
\label{asymptotic corr}
\Sigma_2(t)&=\frac{\Sigma_1(0)}{N}-\frac{4D\tau_c}{N^2(1-r^2)}
+\frac{2D}{N}t~.
\end{align}
\end{subequations}
Since the variance $\Sigma_1(t)$, as well as the two-particle
correlation function $\Sigma_2(t)$, eventually grows linearly in time,
irrespective of the time scale of collisions and the strength of the
driving force, the system does not have a steady state for the
diffusive driving. Moreover, the molecular chaos hypothesis becomes
invalid as the particles in the system becomes more and more
correlated with time.

\section{Conclusion}
\label{conclusion}
In this paper, we have considered a system of Maxwell gas of identical
particles evolving under inelastic binary collisions and external
driving. We illustrated that even though the hierarchy for the evolution of the
distribution functions does not close, those involving the variance
and the two-particle correlation of the velocities close exactly ---
without any approximations.  We also find the exact tail of the velocity distribution in the steady state.
Both the driving and the  interparticle collisions are treated in continuous time as Poisson
processes. The Ornstein-Uhlenbeck
driving (and so also the diffusive driving) can be realized as a special case.
From the exact evolution of
the variance and the two-particle correlation function, the conditions
for a system to be in a steady state can be obtained. Our calculations
show that with the diffusive driving the system cannot have a steady
state as the energy and correlations eventually increase linearly with
time, as found earlier for discrete time dynamics~\cite{prasad}.

\begin{acknowledgements}
 We acknowledge the hospitality of the GGI, Florence, during the
workshop ``Advances in Nonequilbrium Statistical Mechanics'',
 where part of this work was carried out. S.S.
acknowledges the support of the Indo-French Centre for the Promotion
of Advanced Research (IFCPAR/CEFIPRA) under Project 4604-3. A.D. thanks
DST for support through the Swarnajayanti fellowship.
\end{acknowledgements}

\appendix

\section{Maxwell model with discrete time dynamics}
\label{discrete}

In this appendix we briefly review the discrete-time version of the model~\cite{prasad}. 
 The system evolves in discrete time steps as follows. At each time
step, with a probability $p$, a pair of particles (say, $i$ and $j$) is
chosen [of $N (N-1)/2$ pairs] at random and the velocities are
modified from $(v_i^*,v_j^*)$ to $(v_i, v_j)$ according to the rule of
inelastic collision given by \eref{interparticle collision}.  With the
remaining probability $1-p$, a single particle is selected (of $N$
particles) at random and its velocity is modified according
to \eref{wall collision}.

Note that this particular driving scheme differs slightly from the one
employed in Ref.~\cite{prasad} where the forcing was done simultaneously on
two particles independently. However, this does not alter the
qualitative behaviors. Also, unlike the Maxwell model with the rate of
collisions proportional to the root-mean-square velocity at that
time~\cite{Ben-naim:02}, here the probability $p$ is assumed to be
constant over time, as in~ Ref.~\cite{Ben-naim:00}. This would, of course,
change time-dependent behaviors. For example, as discussed
in \aref{HCS} and shown in \fref{Figure2}, in the absence of the
external drive, the mean energy decays exponentially, as opposed to
the Haff's cooling law. However, our main focus here is in the steady-state properties, which are unchanged even if the collision
rates or probabilities are taken to be constant over time; this
makes the analysis relatively simpler.

Let $v_{i}(n)$ be the velocity of the $i$-{\rm th} particle at the
$n$-{\rm th} time step. The variance $\Sigma_1(n)$ and the
two-particle correlation function $\Sigma_2(n)$ are defined as
\begin{subequations}
\label{Sigma}
\begin{align}
 \Sigma_1(n)&= \frac{1}{N}\; \sum^{N}_{i=1}\; \langle
 v^2_{i}(n)\rangle,\\
\Sigma_2(n)&= \frac{1}{N(N-1)}\; \sum_{i\neq j}\; \langle v_{i}(n)
 v_{j}(n)\rangle, 
\end{align}
\end{subequations}
 respectively, with the angular brackets denoting the ensemble
 average. It turns out that their evolution follows an exact recursion
 relation given by
\begin{equation}
\label{discrete_evln_eqn}
 X_n={\bf R}_dX_{n-1}+C_d
\end{equation}
where
\begin{equation*}
X_n=\left[\begin{array}{c}
\Sigma_1(n) \\ \Sigma_2(n)
\end{array}\right] ,\qquad C_d= \left[\begin{array}{c}
(1-p)\frac{\sigma^{2}}{N}\\ 0\end{array}\right], 
\end{equation*}
and
\begin{equation}
{\small{
{\bf R}_d=\left[\begin{array}{cc} 
1- \frac{\left[p(1-r^2)+(1-p)(1-r^{2}_{w})\right]}{N}
& \frac{p(1-r^2)}{N} 
\\[5mm]
\frac{p(1-r^2)}{N(N-1)} & 1-
\frac{\left[p(1-r^2)+
2(N-1)(1-p)(1+r_{w})\right]}{N(N-1)}\end{array}\right]}}
\label{discrete matrix}
\end{equation}

Now, for the case $r_w=-1$, one of the eigenvalues of ${\bf R}_d$ is
unity, resulting in the variance and the two-particle correlation to
eventually increase linearly with number of time-steps $n$. Therefore,
the system does not reach a steady state for this particular case
$r_w=-1$.

Since $\mathbf{R}_d$ is a positive matrix, the Perron-Frobenius
theorem guarantees that it has a real positive eigenvalue
(Perron-Frobenius eigenvalue) such that the other eigenvalue is
strictly less than this, in absolute value. This can be indeed
verified easily for a $2\times 2$ matrix by an explicit calculation.
The other eigenvalue is also real, which also follows from the fact
that the complex eigenvalues of a real matrix must occur in conjugate
pairs.  The Perron-Frobenius eigenvalue is bounded from above (below)
by the maximum (minimum) of the row sums of the matrix.  For $ -1 <
r_w \le 1$, it is immediately evident, from the explicit form of the
above matrix, that both the row sums are less than unity. Thus, both
the eigenvalues are less than unity, in absolute value, and hence the
system reaches a steady state. In the steady state, the variance and
the two-particle correlation function can be found as
\begin{widetext}
\begin{subequations}
\label{steady state discrete}
\begin{align}
\Sigma_1^\mathrm{ss}=\frac{\sigma^2 \bigl[(1-r^2)+4\gamma(1+r_w)(N-1)\bigr]}
{(1-r_w^2)(1-r^2) + 2 (1+r_w) (N-1)\bigl[(1-r^2) + 2\gamma(1-r^2_w)\bigr]}~\\
\Sigma_2^\mathrm{ss}=\frac{\sigma^2(1-r^2)}
{(1-r_w^2)(1-r^2) + 2 (1+r_w) (N-1)\bigl[(1-r^2) + 2\gamma(1-r^2_w)\bigr]}~,
\end{align}
\end{subequations}
\end{widetext}
where $\gamma=(1-p)/(2p)$.  In the $N\to\infty$ limit, the
steady-state variance, $\Sigma_1^\mathrm{ss}$ becomes independent of
$N$,
\begin{equation}
\Sigma_1^\mathrm{ss}=\frac{2\gamma\sigma^2}{(1-r^2)+2\gamma(1-r_w^2)},
\label{steadystate_variance}
\end{equation}
while the two-particle correlation function $\Sigma_2^\mathrm{ss}$
vanishes as $O(N^{-1})$.  Therefore, in the limit of large $N$, the
steady-state single-particle probability distribution closes; the
moment-generating function $Z(\lambda)=\langle \exp(-\lambda
v)\rangle$ of the steady-state velocities can be shown to satisfy the
equation
\begin{equation}
Z(\lambda)=q Z(\epsilon\lambda)Z([1-\epsilon]\lambda) 
+(1-q) Z(r_w \lambda)f(\lambda),
\label{recursion_mgf}
\end{equation}
where $q=2 p/(1+p)$, $\epsilon=(1-r)/2$.  This equation is identical
to \eref{Z-continuous} obtained for the continuous time dynamics.

\section{Homogeneous cooling state}
\label{HCS}

Here, we obtain the freely cooling behavior of the system in the
absence of driving by setting $C=0$, (i.e., $\sigma=0$)
in \eref{continuum matrix abstract finite gamma}, which mimics a
system in a static box. We first consider the linear case $g=1$ and
afterwards consider the case where $g=\Sigma_1^{1/2}$.

\subsection{The linear case: $g=1$}

For $g=1$, the linear equation \eqref{continuum matrix abstract finite
gamma} with $C=0$, can be solved exactly, which gives
\begin{subequations}
\label{eq-1}
\begin{align}
&\Sigma_1(t)=\frac{\Sigma_1(0)}{\lambda_+
-\lambda_-} \left[(R_{22}-\lambda_-) e^{-\lambda_- t} +
(\lambda_+-R_{22}) e^{-\lambda_+ t} \right], \\ 
&\Sigma_2(t)=\frac{R_{21}\Sigma_1(0)}{\lambda_+
-\lambda_-}\left[e^{-\lambda_- t} -e^{-\lambda_+ t} \right],
\end{align}
\end{subequations}
where $R_{ij} = |\mathbf{R}_{ij}|$ denote the absolute values of the
elements of the matrix $\mathbf{R}$ given by \eref{collision wall
dissipation point process} and $-\lambda_\pm$ are the eigenvalues of
the matrix $\mathbf{R}$, given by
\begin{align}
\label{lambda-pm}
\lambda_\pm &= \frac{1}{2}\left[(R_{11} + R_{22}) \pm
\sqrt{(R_{11} + R_{22})^2 -4 (R_{11} R_{22} -R_{12} R_{21}}
 \right]\notag\\
&=\frac{1}{2}\left[(R_{11} + R_{22}) \pm
\sqrt{(R_{11} -R_{22})^2 +4 R_{12} R_{21}}
 \right].
\end{align}
 Note that $\lambda_\pm > 0$ for $-1 < r_w \le 1$. In \fref{Figure2}, we plot
$\Sigma_{1,2}$ as a function of $t$, as given by \eref{eq-1}.

\subsection{The non-linear case: $g=\Sigma_1^{1/2}$}
In this case, $\Sigma_1$ and $\Sigma_2$ evolve by
\begin{subequations}
\label{eq0}
\begin{align}
\label{eq1}
\frac{d\Sigma_1}{dt} &= -R_{11} \Sigma_1^{3/2} +
R_{12} \Sigma_1^{1/2}\Sigma_2, \\
\label{eq2}
\frac{d\Sigma_2}{dt} &= R_{21} \Sigma_1^{3/2} -
R_{22} \Sigma_1^{1/2}\Sigma_2.
\end{align}
\end{subequations}
\Eref{eq2} for $\Sigma_2$ can be solved exactly in terms of
$\Sigma_1$ as
\begin{equation}
\Sigma_2(t)= R_{21}\int_0^t dt_1\Sigma_1^{3/2}(t_1) 
\exp \left[-R_{22}\int_{t_1}^t\Sigma_1^{1/2}(t_2) \,dt_2 \right],
\label{eq_sigma2_int}
\end{equation}
where we have used the initial condition $\Sigma_2(0)=0$. On the other
hand, it is difficult to obtain an exact solution of \eref{eq1} for
$\Sigma_1$.  Nevertheless, near $t=0$, using the initial condition
$\Sigma_2(0)=0$, we can write \eref{eq1} as $d\Sigma_1/dt\approx
-R_{11}\Sigma_1^{3/2}$, which yields
\begin{equation}
\Sigma_1(t) \approx \frac{\Sigma_1(0)}{\left(1+\frac{1}{2}
R_{11} \Sigma_1^{1/2}(0)\, t \right)^2}.
\label{eq3}
\end{equation}
Now, substituting the above expression for $\Sigma_1(t)$
in \eref{eq_sigma2_int}, after carrying out the integration, we obtain
\begin{equation}
\Sigma_2(t)\approx \frac{R_{21}/R_{11}}{1-\theta}
\left[\frac{\Sigma_1(0)}{\left(1+\frac{1}{2}
R_{11} \Sigma_1^{1/2}(0)\, t \right)^{2\theta}}- \Sigma_1(t) \right],
\label{eq4}
\end{equation}
for $\theta\not=1$, where $\theta= R_{22}/R_{11}$. For $\theta=1$, we
get
\begin{equation}
\Sigma_2(t)\approx 
\frac{2({R_{21}/R_{11}})\,\Sigma_1(0)\,\ln \left(1+\frac{1}{2}
R_{11} \Sigma_1^{1/2}(0)\, t \right)}{\left(1+\frac{1}{2}
R_{11} \Sigma_1^{1/2}(0)\, t \right)^{2}}. 
\end{equation}
Therefore, for large $t$, we have $\Sigma_2(t) \sim t^{-2\theta}$ for
$\theta <1$, whereas $\Sigma_2(t) \sim t^{-2}$ for $\theta >1$. For
$\theta=1$, there is a logarithmic correction $\Sigma_2(t) \sim (\ln
t) t^{-2}$.

Now, if we take $\tau_c$ to be proportional to the total number of
pairs and $\tau_w$ to be proportional to the number particles, then
for large $N$, we have $\tau_c$ is $O(N^{-2})$ and $\tau_w$ is
$O(N^{-1})$. Consequently, we see from \eref{collision wall
dissipation point process} that $R_{11}$, $R_{12}$, and $R_{22}$ are
$O(N^{-1})$, whereas $R_{21}$ is $O(N^{-2})$. Therefore, the prefactor
outside the square bracket in the expression \eqref{eq4} is
$O(N^{-1})$, and hence the second term on the right-hand side
of \eref{eq1} can be neglected even beyond the small $t$ region, for
large $N$.  As a result, the expression \eqref{eq3} and
hence \eref{eq4} remain valid even for large times. Essentially, for
the freely cooling gas, the two-particle correlation is not important.
In the limit $N\to\infty$, the exponent $\theta$ is given by
\begin{equation}
\theta=\frac{4\gamma(1+r_w)}{1-r^2 + 2 \gamma(1-r_w^2)}. 
\end{equation}
\Fref{Figure2} compares the expressions \eqref{eq3} and \eqref{eq4} with the
exact values obtained by numerically solving \eref{eq0}.

\end{document}